\documentclass[%
mathleft,%
final,%
]{an}
\pdfoutput=1
\usepackage[pdftex]{graphicx}
\usepackage{times}
\usepackage{natbib}
\bibpunct{(}{)}{;}{a}{}{,}
\def\apjl{ApJ}%
\def\prd{Phys.~Rev.~D}%
\newcommand{\ms}{m\,s$^{-1}$}
\newcommand{\cms}{cm\,s$^{-1}$}
\begin{document}

% The following seven commands are intended for editorial usage and
% should be ignored by the author(s).
\Pagespan{10}{}% Document's page range.
% If second parameter is left empty, the last page is computed
% automatically.
\Yearpublication{2014}%
\Yearsubmission{2013}%
\Month{8}%
\Volume{335}%
\Issue{1}%
\DOI{12.1002/asna.201312004}%

\title{ESPRESSO: The next European exoplanet hunter}
\author{F. Pepe\inst{1}\fnmsep\thanks{Corresponding author:
  {molaro@oats.inaf.it}}
 \and P. Molaro  \inst{2,4}
\and S. Cristiani\inst{2}
\and R. Rebolo\inst{3}
\and N.C. Santos\inst{4,12}
\and H. Dekker \inst{5}
\and D. M\'egevand \inst{1}
\and F.M. Zerbi \inst{6}
\and A. Cabral \inst{7}
\and P. Di Marcantonio \inst{2}
\and M. Abreu  \inst{7}
\and M. Affolter \inst{9}
\and M. Aliverti \inst{6}
\and C. Allende Prieto \inst{3}
\and M. Amate \inst{3} 
\and G. Avila \inst{5} 
\and V. Baldini \inst{2}
\and P. Bristow \inst{5}
\and C. Broeg \inst{9}
\and R. Cirami \inst{2}
\and J. Coelho \inst{7}
\and P. Conconi \inst{6}
\and I. Coretti \inst{2}
\and G. Cupani \inst{2}
\and V. D'Odorico \inst{2}
\and V. De Caprio \inst{6}
\and B. Delabre \inst{5} 
\and R. Dorn \inst{5}
\and P. Figueira \inst{4}
\and A. Fragoso \inst{3}
\and S. Galeotta \inst{2}
\and L. Genolet \inst{1}  
\and R. Gomes \inst{8} 
\and J.I. Gonz\'alez Hern\'andez \inst{3}
\and I. Hughes \inst{1}
\and O. Iwert \inst{5}
\and F. Kerber \inst{5}
\and M. Landoni \inst{6}
\and J.-L. Lizon \inst{5}
\and C. Lovis \inst{1}
\and C. Maire \inst{1}
\and M. Mannetta \inst{6}
\and C. Martins \inst{4}
\and M. Monteiro \inst{4}
\and A. Oliveira \inst{8}
\and E. Poretti \inst{6}
\and J.L. Rasilla \inst{3}
\and M. Riva \inst{6}
\and S. Santana Tschudi \inst{3}
\and P. Santos \inst{8}
\and D. Sosnowska \inst{1}
\and S. Sousa \inst{4}
\and P. Span\'o \inst{11}
\and F. Tenegi \inst{3}
\and G. Toso \inst{6}
\and E. Vanzella  \inst{2} 
\and M. Viel \inst{2}
\and M.R. Zapatero Osorio \inst{10}
}
\titlerunning{ESPRESSO spectrograph}
\authorrunning{F. Pepe et al.}
\institute{
Observatoire de l'Universit\'e de Gen\'eve, Ch. des Maillettes 51, CH-1290 Versoix, Switzerland
 \and INAF -- Osservatorio Astronomico di Trieste, Via Tiepolo 11, I-34143 Trieste, Italy
\and Instituto de Astrofisica de Canarias,  Via Lactea, E-38200 La Laguna, Tenerife, Spain
\and Centro de Astrofisica da Universidade do Porto, Rua das Estrelas, 4150-762 Porto, Portugal
\and ESO, European Southern Observatory, Karl-Schwarzschild-Stra\ss{}e 2, 85748 Garching, Germany 
\and
INAF -- Osservatorio Astronomico di Brera, Via Bianchi 46, I-23807 Merate, Italy
\and CAAUL, Faculty of Sciences, Univ. of Lisbon, Tapada da Ajuda, Edificio Leste, 1349-018 Lisbon, Portugal
\and LOLS, Faculty of Sciences, Univ. of Lisbon, Estrada do Paco do Lumiar 22, 1649-038 Lisbon, Portugal
\and Physics Institute of University of Bern, Siedlerstra\ss{}e 5, CH-3012 Bern, Switzerland
\and Centro de Astrobiolog\'ia, INTA, Carrettera Ajalvir km 4, 28850, Torrej\'on de Ardoz, Madrid, Spain
\and NRCC-HIA, 5071 West Saanich Road Building VIC-10, Victoria, British Columbia V9E 2E, Canada
\and Departamento de Física e Astronomia, Faculdade de Ciências, Universidade do Porto, Rua das Estrelas, 4150-762 Porto, Portugal
 }

\received{2013 Aug 29}
\accepted{2013 Nov 1}
\publonline{2014 Jan 15}
%\keywords{cosmology: observations -- quasars: absorption lines -- atomic processes -- line: formation}
\keywords{instrumentation: spectrographs -- plantary systems -- techniques: spectroscopic}
\abstract{%
The acronym ESPRESSO stems for {E}chelle {SP}ectrograph for {R}ocky {E}xoplanets and {S}table {S}pectroscopic {O}bservations; this instrument  will be the  next VLT high resolution  spectrograph. The spectrograph will be installed at the Combined-Coud\'e Laboratory of the VLT and  linked to the four 8.2\,m Unit Telescopes (UT) through four optical Coud\'e trains. ESPRESSO   will combine  efficiency  and  extreme spectroscopic precision.  ESPRESSO is foreseen  to achieve a gain of two magnitudes with respect to its predecessor HARPS,  and to improve the  instrumental radial-velocity precision to reach the 10 \cms~ level.  It can be operated either with a single UT or with up to four UTs, enabling an additional gain in the latter mode. The incoherent combination of four telescopes and the extreme precision requirements called for many innovative design solutions while ensuring the technical heritage of the successful HARPS experience.
 ESPRESSO will allow to explore new frontiers in most domains of astrophysics that  require precision and sensitivity. The main scientific drivers are the search and characterization of rocky exoplanets in the habitable zone of quiet, nearby G to M-dwarfs and the analysis of the variability of fundamental physical constants. 
The project  passed the final design review in May 2013 and  entered the manufacturing phase. ESPRESSO will be installed at the Paranal Observatory in 2016 and its operation  is planned to start  by the end of the same year.
}
\maketitle
\sloppy

\section{Introduction}
 High-resolution spectroscopy provides  physical insights in the study of  stars, galaxies, and interstellar and intergalactic medium.    Besides the importance of  observing fainter and fainter objects by  increasing the  photon collecting area  by making bigger  telescopes,  the importance  of high-precision  has emerged  in recent years as a crucial element in spectroscopy. In many investigations   repeatable observations over long temporal  baseline  are needed.    For instance,  the HARPS spectrograph at the ESO 3.6-m telescope  is  a pioneering instrument   for precise radial-velocity (RV) measurements \citep{may03}.  The search for terrestrial planets in habitable zone is one of the most exciting science  topics of the next decades  and  one of the main  drivers for the new generation of Extremely Large Telescopes. 
 The need for a similar instrument    on the VLT    has been emphasized  in the ESO-ESA working group report on extrasolar planets. 
In October 2007  the  ESO STC recommended  the development of additional second-generation VLT instruments, and this  proposal was endorsed by the ESO Council   in December of the same year. Among the recommended instruments, a high-resolution, ultra-stable spectrograph for the VLT combined-Coud\'e  focus arose as a cornerstone to complete the current 2nd generation VLT instrument suite. Following these recommendations,  in March 2008 ESO issued a call for proposals   to member state institutes or consortia  to carry out the Phase-A study for such a instrument. The submitted proposal was accepted by ESO and the ESPRESSO Consortium% 
   \footnote{The ESPRESSO Consortium is composed of: 	Observatoire Astronomique de l'Universit\'e de Gen\'eve (project head, Switzerland); 	Centro de Astrof\'isica da Universidade do Porto (Portugal),	Faculdade de Ciencias da Universidade de Lisboa (Portugal); INAF-Osservatorio Astronomico di Brera (Italy);	INAF-Osservatorio Astronomico di Trieste (Italy); Instituto de Astrof\'isica de Canarias (Spain);	Physikalisches Institut der Universit\"at Bern (Switzerland;) ESO participates to the ESPRESSO project as Associated Partner and contributes the Echelle grating, the camera lenses, the detector system and the cryogenic and vacuum control system.} was selected to carry out   the construction of this spectrograph  in collaboration with ESO. 
The  project  kick-off was held in February 2011 and the design phase  ended with the Final Design Review  in May 2013. The procurement of components and manufacturing of subsystems will last about 18 months and  early 2015 the subsystems will be ready for integration in Europe. Acceptance Europe of the instrument will be held in the  fall 2015 while  the transfer, installation and commissioning of the instrument at  Paranal will take place in 2016. Acceptance Paranal is planned to take place at the end of 2016.

\section{What science?}
The main scientific drivers for the new spectrograph   were defined by ESO as follows:
 \begin{itemize}
\item{	measure high-precision RV for search for rocky planets;}
\item{	measure the variation of physical constants;}\item{	analyse the chemical composition of stars in nearby galaxies.}
\end{itemize}

 \begin{figure}
 \vskip-2mm
\includegraphics[width=83mm]{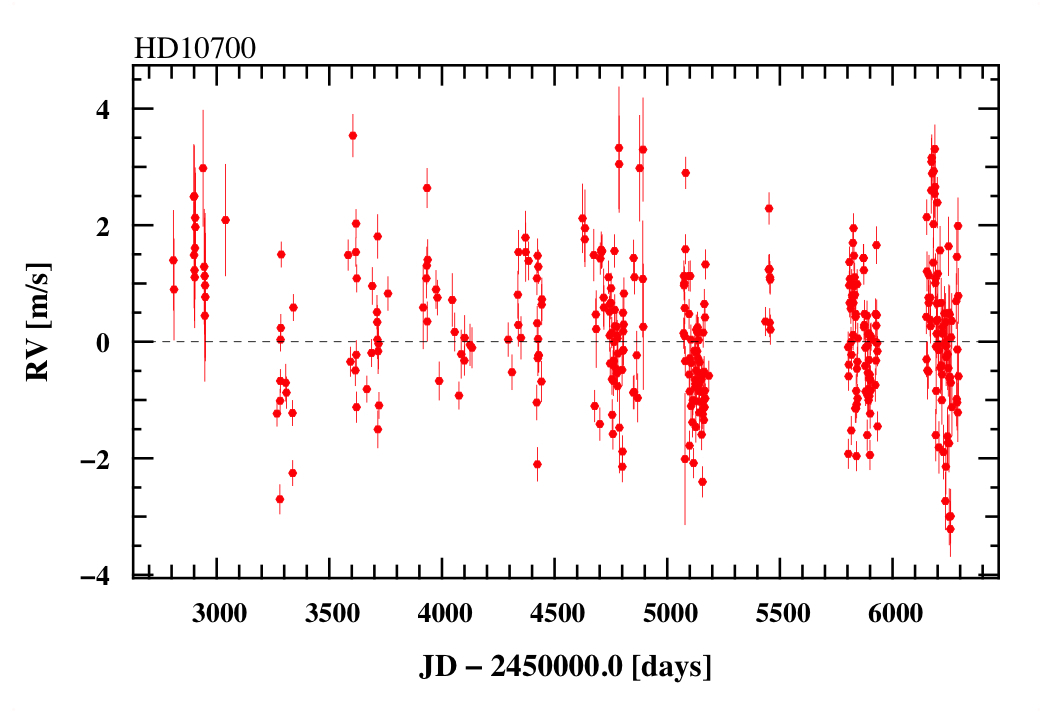}
\caption{  Ten years of $\tau$ Ceti's RVs as measured by HARPS. The overall dispersion is of 1 \ms .  
}
\label{label11}
\end{figure}

% \begin{figure}
%\includegraphics[width=50mm,height=28mm]{Fig1.jpg}
%\includegraphics[width=80mm]{fig13.jpg}
%\caption{  Phase-folded RV variation induced by the P = 58-days period around the K5 dwarf HD 85512.}
%\label{label13}
%\end{figure}

\subsection{Searching for rocky planets in the habitable zone}

Terrestrial planets in the habitable zone of their parent stars are one of the main scientific topics of the next decades in Astronomy, and one of the main science drivers for the new generation of extremely large telescopes. ESPRESSO, being capable of achieving a precision of 10 \cms\ in terms of RV, will be able to register the signals of Earth-like planets in the habitable zones (i.e., in orbits where water is retained in liquid form on the planet surface) around nearby solar-type stars and around stars smaller than the Sun. Since 1995  research teams  using the RV technique  have discovered about 600 extrasolar planets, some of which with only a few times the mass of the Earth. Today, dozens of detected RV extrasolar planets have estimated masses  below 10 Earth masses, and most of them were identified using the HARPS spectrograph (e.g. Mayor et al. 2011). The rate of these discoveries increases steadily. The HARPS high-precision RV program has shown that half of the solar-like stars in the sky harbour Neptune-mass planets and super-Earths, a finding also supported by the recent discoveries of the {\it Kepler} satellite (e.g. Howard et al. 2012). These exciting discoveries were made possible thanks to the sub-m\,s$^{-1}$ precision reached by HARPS. In  Fig. \ref{label11}  are shown 10 years of  RVs  measured by HARPS for Tau Ceti. The overall dispersion is of 1 \ms, but  time-binning of the data will reduce the dispersion  down to 20 \cms\ with  the absence of any long-term trend. 
 Most of the discovered exoplanets  would have remained out of reach if  existing facilities  had been  limited to 3 \ms. 

Considering the observational bias towards large masses,  one should expect a huge amount of still undiscovered low-mass planets, even in already observed stellar samples. The most recent planet formation  theories  support this   view  and  the detected   population is probably only the tip of the iceberg, and    the bulk is  still  to be discovered. ESPRESSO is designed to explore this new mass domain and charter unknown territories. This goal can only be obtained by combining high efficiency with high instrumental precision. ESPRESSO will be optimized to obtain best RVs on quiet solar-type stars. A careful selection of these stars will allow to focus the observations on the best-suited candidates: non-active, non-rotating, quiet G to M dwarfs.  An optimized observational strategy will permit the characterization of the planetary systems and very low-mass planets despite stellar noise. An impressive demonstration that this approach is realistic has been recently delivered   by detecting   a 1 Earth-mass planet around our neighbour $\alpha$\,Cen~B \citep{dum12}.  
%Another  example  is given by the discovery of HD 85512 b, a 3.6 earth-mass planets just at the edge of the habitable zone of a K5 dwarf (Fig \ref{label13}).

 \begin{figure}
 \vskip-2mm
\includegraphics[width=83mm]{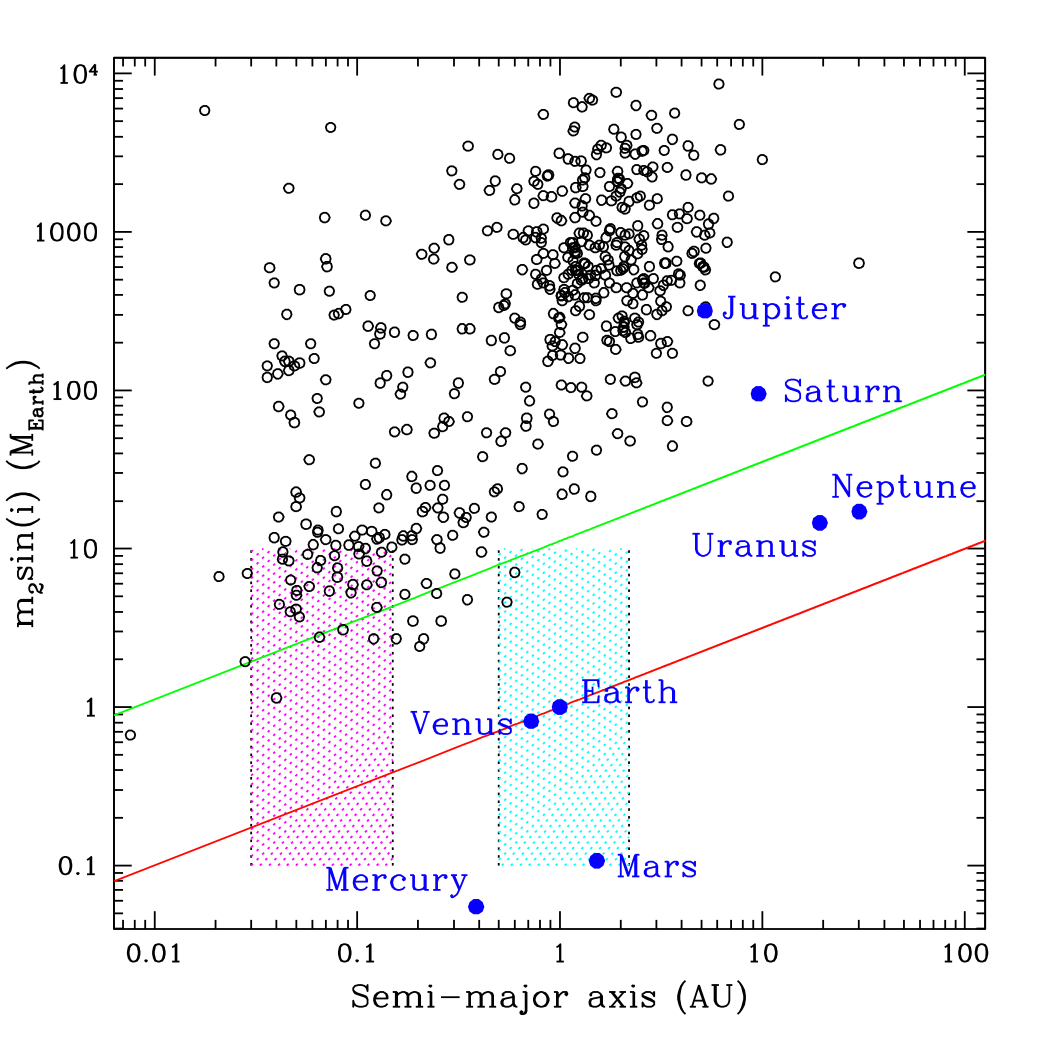}
\caption{  Detectability of planets orbiting a 0.8 M$_{\odot}$ star (red solid line) and a 1 M$_{\odot}$ star (green solid line) in the mass vs. semi-major axis plane expected for ESPRESSO. The detectability curves have been calculated assuming a velocity amplitude of 10 cm\,s$^{-1}$  (for the 0.8 M$_{\odot}$ star) and 1 \ms\ (for the 1 M$_{\odot}$ star), null eccentricity, and $\sin i$ = 1. Known RV planets of solar-type stars are plotted as open circles, and the planets of the solar system (solid circles) are labeled. The ``habitable zones" of 0.8--1.2 M$_{\odot}$ and 0.2--0.3  M$_{\odot}$  stars are indicated with blue and pink dotted lines, respectively. These are regions where rocky planets with a mass in the interval 0.1--10 of the mass of the Earth    can retain liquid water on their surface.  \vspace{-2mm}
}
\label{label12}
\end{figure}

With a precision of 10 \cms\ (about a factor of 10 better than HARPS), it will
be possible to detect rocky planets down to the Earth mass in the habitable zone
of solar-type stars. For comparison, the Earth imposes a velocity amplitude of 9
\cms\ onto the Sun.  As shown in  Fig. \ref{label12} by extending the sample towards the lighter M-stars, the task becomes even easier since the RV signal increases with decreasing  stellar mass.  ESPRESSO will operate at the peak of its efficiency for a spectral type up to M4-type stars.  The discovery and the characterization of a new population of very light planets will open the door to a better understanding of planet formation and deliver new candidates for follow-up studies by using other techniques such as  transit, astrometry, Rossiter-McLaughlin effect, etc.
Another important task for ESPRESSO will be the follow-up of transiting planets. It should be recalled that many \emph{Kepler} transit candidates are too faint to be confirmed by existing RV instruments.  Other satellites like GAIA, TESS and
 hopefully PLATO will provide us with many new transit candidates, possibly
 hosted by bright stars. ESPRESSO will be the ideal (and maybe unique) machine
 to make spectroscopic follow-up of Earth-size planets discovered by the transit
 technique. Besides, being an exquisite RV machine, ESPRESSO will provide
 extra-ordinary and stable spectroscopic observations, opening new possibilities
 for transit spectroscopy and analysis of the light reflected by the exoplanet.
 Several groups are currently investigating to which extent this will be
 feasible in the visible and infrared spectral domain (see, e.g., Snellen 2013;
 Snellen et al. 2013, and references therein). Fast-cadence spectra of the most promising candidates will provide estimates of the maximum frequency of solar-like oscillations. The resulting seismic constraints on the gravity of the host stars and precise spectroscopic analysis will allow us to improve the determination of the mass and radius of the star and, therefore, of the planet (Chaplin \& Miglio 2013).
ESPRESSO should  be considered  an important  development step towards  obtaining high-precision spectrographs on the  ELT. 

\subsection{Do physical constants vary?}
The standard model of particle physics depends on many ($\approx$\,27)
independent numerical parameters that determine the strengths of the different
forces and the relative masses of all known fundamental particles. There is no
theoretical explanation for their actual value, but  nevertheless they determine
the properties of atoms, cells, stars and the whole Universe. They are commonly
referred to as the fundamental constants of nature, although most of the modern
extensions of the standard model predict a variation of these constants at some
level (see Uzan 2011; Bonifacio et al. 2013). For instance, in any theory involving more than four space-time dimensions, the constants we observe are merely four-dimensional shadows of the truly fundamental high dimensional constants. The four dimensional constants will then be seen to vary as the extra dimensions change slowly in size during their cosmological evolution. An attractive implication of quintessence models for the dark energy is that the rolling scalar field produces  a negative pressure and therefore the acceleration of the universe may couple with other fields and be revealed by a change in the fundamental constants (Amendola et al. 2013). Earth-based laboratories have so far revealed no variation in their values. For example, the constancy of the fine structure constant $\alpha$ is ensured to within a few parts per 10$^{-17}$  over a 1-yr period (Rosenband et al. 2008). Hence their status as truly {\it constants} is amply justified.  

\begin{figure}
 \vskip-2mm
\includegraphics[width=83mm]{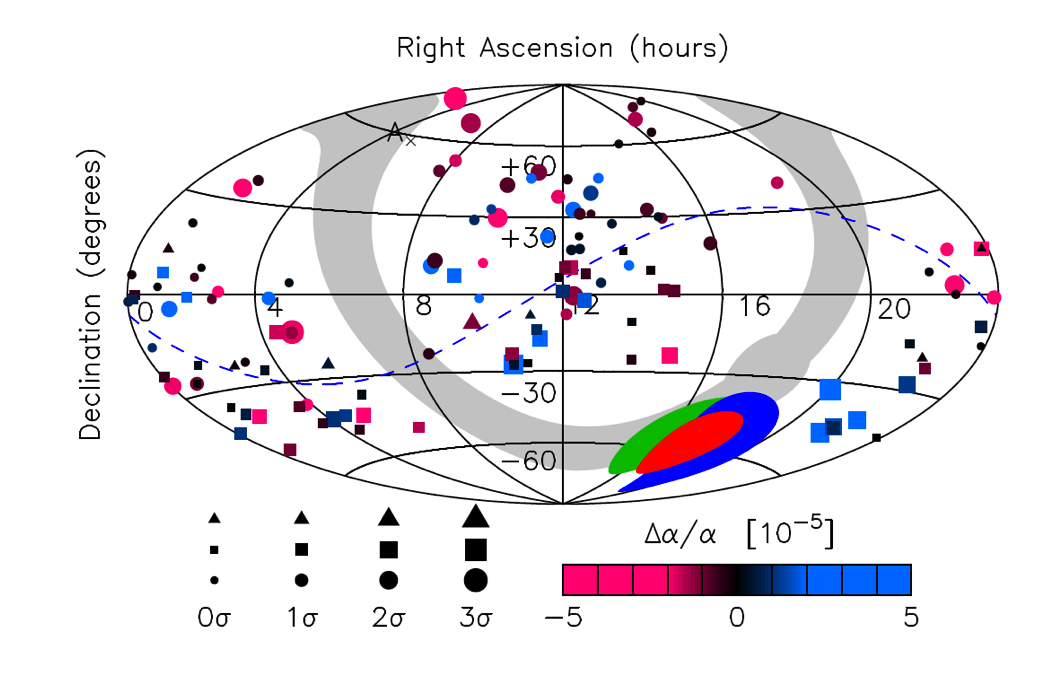}
\caption{  All-sky spatial dipole with the combined VLT (squares) and Keck (circles) $\alpha$ measurements from Webb et al. (2011). Triangles are measures in common with  the two telescopes. The blue dashed line shows the equatorial region of the dipole.
}
\label{fig14}
\end{figure}

Astronomy has a great potential in probing their variability at very large distances and in the early Universe. In fact, the transition frequencies of the narrow metal absorption lines observed in the spectra of distant quasars are sensitive to $\alpha$ (e.g. Bahcall et al. 1967) and those of the rare molecular hydrogen clouds are sensitive to $\mu$, the proton-to-electron mass ratio (e.g. Thompson 1975).
With the advent of 10-m class telescopes, observations of spectral lines in distant QSOs gave the first hints that the value of the fine structure constant might change over time, being lower in the past by about 6 ppm (Webb et al. 1999; Murphy et al. 2004). The addition of other 143 VLT/UVES absorbers (Fig. \ref{fig14}) have revealed a 4$\sigma$  evidence for a dipole-like variation in $\alpha$ across the sky at the 10 ppm level (Webb et al. 2011; King 2012). Several other constraints from higher-quality spectra of individual absorbers exist but none of them  directly supports or strongly conflicts with the $\alpha$ dipole evidence, and a possible systematic producing opposite values in the two hemispheres is not easy to identify.

In order to probe $\mu$,  the H$_2$ absorbers need to be at a redshift $z$ $>$ 2--2.5 to place the Lyman and Werner H$_2$ transitions redwards of the atmospheric cut-off. Only five systems have been studied so far, with no current indication of variability at the level of ~10 ppm (e.g. Rahmani et al. 2013). At lower redshifts precise constraints on $\mu$-variation are available from radio- and millimeter-wave spectra of cool clouds containing complex molecules like ammonia and methanol (see, e.g., Flambaum \& Kozlov 2007; Levskakov et al. 2013). Other techniques involving radio spectra typically constrain combinations of constants by comparing different types of transitions (e.g. electronic, hyperfine, rotational, etc.).

Extraordinary claims require extraordinary evidence, and a confirmation of variability with high statistical significance is of crucial importance. Only a high resolution spectrograph that combines a large collecting area with extreme wavelength precision can provide definitive clarification. A relative variation in $\alpha$ or $\mu$ of 1 ppm leads to velocity shifts of about 20 \ms\ between typical combinations of transitions. ESPRESSO is expected to provide an increase in the accuracy of the measurement of these two constants by at least one order of magnitude compared to VLT/UVES or Keck/HIRES. More stringent bounds are also important and the ones already provided  constrain  the parameter space  of various theoretical models that predict their variability.

\subsubsection{Chemical composition of stars in local galaxies}

One important piece of information in the understanding of the galaxy formation is the chemical composition of local galaxies. In spite of the many successes in this field, the majority of local galaxies still lack detailed abundance information. For about a dozen of nearby galaxies observable from Paranal  chemical information is available, albeit, except Sagittarius,  for a few stars and for a limited set of elements, and for the faintest galaxies it is based on low to medium resolution spectra. The next nearest galaxy, Leo T, has a distance modulus of 23.1, i.e. more than one magnitude more distant than Leo I (${m-M=21.99}$). The local group galaxies all possess giant stars of magnitude ${V=20}$ or fainter. Although some work has been done with UVES at VLT at this magnitude, it is really difficult   to obtain accurate chemical abundances. For galaxies that possess a young population, like Phoenix or WLM, one can rely on bright O and B supergiants. However, if one considers old metal-poor systems, like Bo\"otes or Hercules, one has to rely on red giants. Although it is clear that most of the chemical information for local galaxies will have to come largely from the ELT, ESPRESSO will give us the chance to have a first but important glimpse into this.

\subsection{A scientific Pandora box}

ESPRESSO combines an unprecedented RV and spectroscopic precision with the largest photon collecting area available today at ESO and unique resolving power (${R \sim 200\,000}$). It will certainly provide breakthroughs in many areas of astronomical research, many of which cannot be anticipated. We shall provide  just few examples below.

\begin{itemize}
\item{{\it The expanding Universe.}
Sandage (1962) first argued that in any cosmological model the redshifts of cosmologically distant objects drift slowly with time. If observed, their redshift drift-rate, ${\rm d}z/{\rm d}t$, would constitute evidence of the Hubble flow's deceleration or acceleration between redshift $z$ and today. Indeed, this observation would offer a direct, non-geometric, completely model-independent measurement of the Universe's expansion history (Liske et al. 2008). ESPRESSO, even in the 4UT mode, is probably not sufficiently sensitive  to measure the tiny signal  which is at the level of few \cms\,yr$^{-1}$. However, it might provide first accurate historical reference measurements and   in any case will represent an important step forward in setting the scene for the next generation of high resolution spectrographs at the E-ELTs. }

 \begin{figure*}
\center{
\includegraphics[width=160mm]{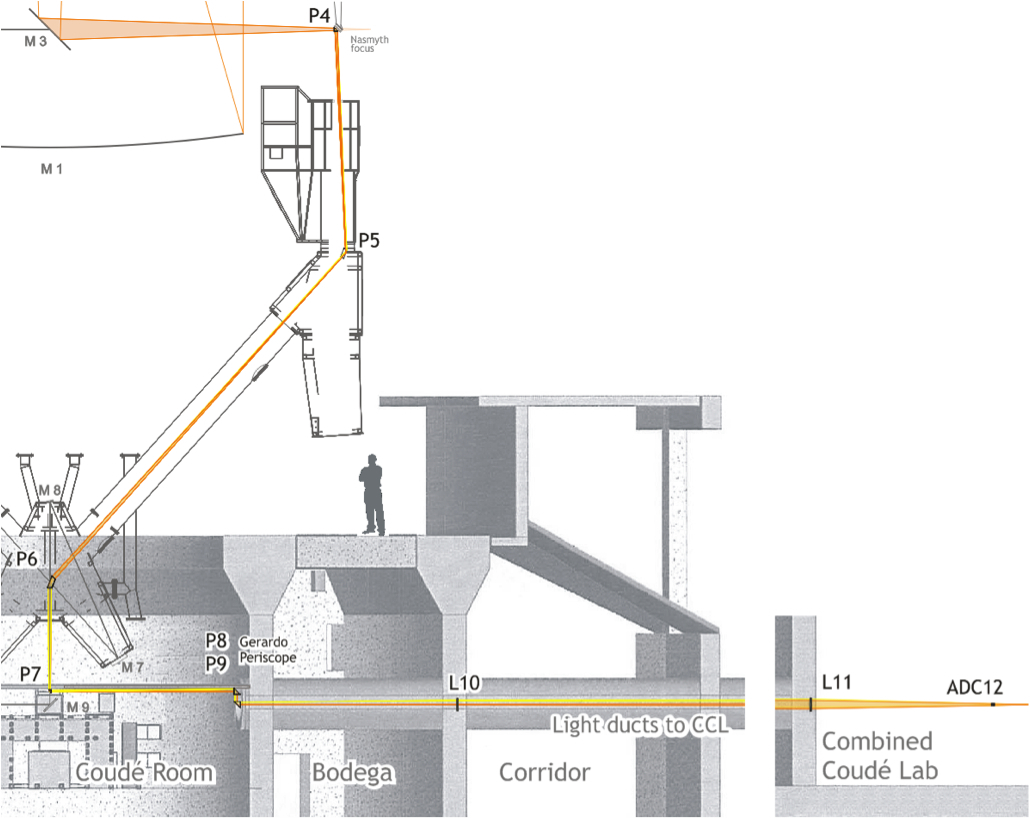}}
\caption{ Coud\'e train of ESPRESSO and optical path through the telescope and the tunnels.
}
\label{label7}
\end{figure*}

\item{{\it Metal poor stars.} The most metal poor stars in the Galaxy are probably the most ancient fossil records of the chemical composition and thus can provide clues on the pre-Galactic phases and on the stars which synthesized the first metals. Masses and yields of Pop.\,III stars can be inferred from the observed elemental ratios in the most metal poor stars (Heger \& Woosley 2010). One crucial question to be answered is the presence of Pop. III low-mass stars. For a long time the Pop. III stars have been thought to be only very massive but the recent discovery of a very metal-poor star with [Fe/H] ${\sim -5.0}$ and {\it normal}  C and N have shown an entirely new picture (Caffau et al. 2011).  Several surveys searching for metal poor stars are currently going on or are planned, and thousands of EMP stars with [Fe/H] ${< -3.0}$, of which maybe several down to [Fe/H]  ${\approx -5.0}$ and hopefully lower, are expected to be found. These will be within the reach of ESPRESSO, which will be able  to provide spectra for an exquisite chemical analysis in both the 1-UT and 4-UT modes.}

\item{{\it Stellar oscillations, asteroseismology, variability.} Stars located in the upper main sequence show non-radial pulsations that cause strong line profile variations. Astroseismic study (i.e., mode identification) of these pulsating stars ($\gamma$ Dor, $\delta$ Sct, $\beta$ Cep, SPB, etc.) provides constraints on the structure of massive stars (e.g. internal convection, overshooting, core size, extension of acoustic and gravity cavities, mass loss phenomena, interplay between rotation and pulsation). ESPRESSO will allow us to perform the short exposures required to identify the high-frequency modes, currently achievable on a wide variety of stars only with photometry from space.}

\item{{\it Galactic winds and tomography of the IGM.}
In principle, spectroscopy of close, multiple, high-redshift quasars allows  to recover the 3-dimensional distribution of matter from the analysis of the H\,{\sc i} Ly$\alpha$ absorption lines. If the multiple lines of sight cross a region where   high-redshift galaxies are present, it is also possible to investigate the properties of outflows and inflows, studying the spectral absorption lines at the redshift of the galaxies and how they evolve moving closer to or far away from the galaxies themselves. The main limitation to the full exploitation of the so-called {\it tomography} of the IGM is the dearth of quasar pairs at the desired separation, bright enough to be observed with the present high-resolution spectrographs at 10\,m-class telescopes. ESPRESSO used in the 4UT mode  would result in a gain of $\approx$\,1.5 magnitude fainter than, e.g.,  UVES, translating into an almost a 20-fold increase  in the number of  observable quasar pairs with separation lower  than 3 arcmin and emission redshift in the range ${2 < z < 3}$.}
\end{itemize}

 \begin{figure}
\includegraphics[width=80mm]{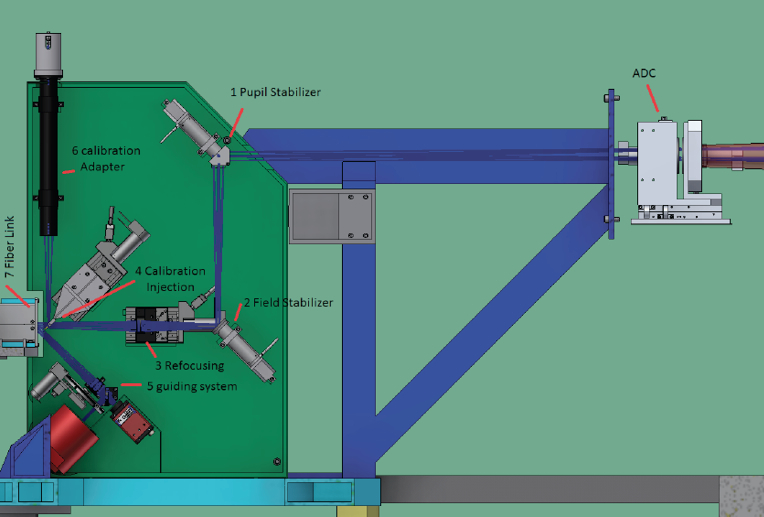}
\caption{  Side view of the Front-End and the arrival of one  UT beam at the CCL. The same is replicated for the other UTs.
}
\label{label8}
\end{figure}

\section{A new-generation instrument for the VLT}

 \begin{figure}
\includegraphics[width=83mm]{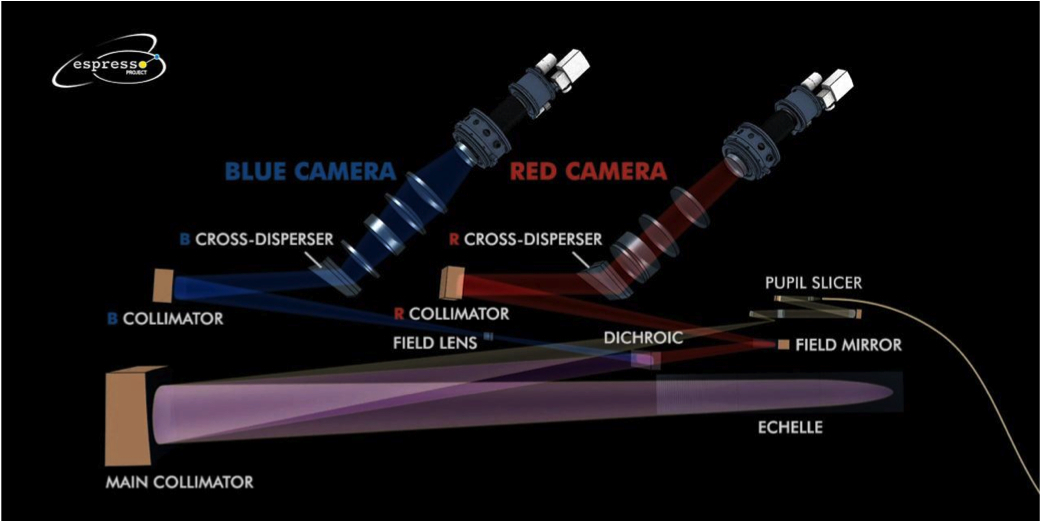}
\caption{ Layout of the ESPRESSO spectrograph and its optical elements.
}
\label{label1}
\end{figure}

ESPRESSO is a fibre-fed, cross-dispersed, high-resolution  \'echelle spectrograph which will be  located in the Combined Coud\'e Laboratory (CCL) at the incoherent focus, where a front-end unit can combine the light from up to 4 Unit Telescopes (UT) of the VLT. The telescope light is fed to the instrument via a so-called Coud\'e-train optical system.  The target and sky light enter the instrument simultaneously through two separate fibres, which  together form the {\it slit} of the spectrograph. 
 Although foreseen since 1977 in the original VLT plan, the incoherent combined focus of the VLT  has never been implemented. Only provision for it, in terms of space left in the UTs structures and ducts in the rock of the mountain, is what is actually available at the VLT. As part of the project agreement, the ESPRESSO Consortium has been asked to materialize such a focus providing the necessary hardware and software as part of the deliverables. The implementation of the Coud\'e train leads to  substantial changes in the Paranal Observatory infrastructure and requires an elaborate interfaces management.  ESPRESSO will be located in the VLT's CCL and, unlike any other instrument built so far, will receive light from any of the four UTs. The light of the single UT scheduled to work with ESPRESSO is then fed into the spectrograph (1-UT mode). Alternatively, the combined light of all the UTs can be fed into ESPRESSO simultaneously (4-UT mode).

 \subsection{The Coud\'e train}
 
 Distances between the UTs and the CCL range between  48m for  UT2 to  69 m for UT1. A trade-off analysis between  solutions based on mirrors, prisms, lenses and  fibres  pointed towards a full optics solution, i.e. using only conventional optics without  fibres for transporting the light from the telescope into the CCL.  The chosen design with the position of the 11 optical elements is shown in Fig \ref{label7}. The Coud\'e train picks up the light with a prism at the level of the Nasmyth-B platform and routes the beam through the UT mechanical structure down to the UT Coud\'e room, and farther to the CCL along the existing incoherent light ducts. The four trains relay a field of 17 arcsec around the acquired object to the CCL. The selected concept to convey the light of the telescope from the Nasmyth focus (B) to the entrance of the tunnel in the Coud\'e room (CR) below each UT unit is based on a set of 6 prisms (with some power). The light is directed from the UT's Coud\'e room towards the CCL using 2 large  lenses. The beams from the four UTs meet  in the CCL, where mode selection and beam conditioning is performed by the  fore-optics of the Front-End subsystem.

 \begin{figure*}%[!hb]
\center{
\includegraphics[width=160mm]{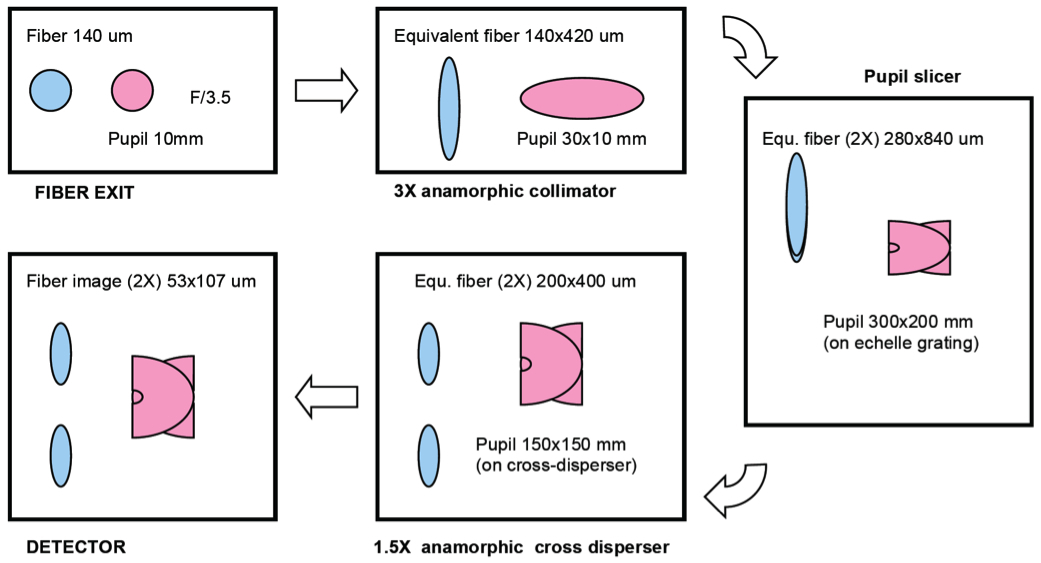}}
\caption{Conceptual description of pupil and fibre image at relevant locations of the spectrograph optical path.
}
\label{label3}
\end{figure*}

\subsection{The Front-End}

\begin{figure}
\includegraphics[width=83mm]{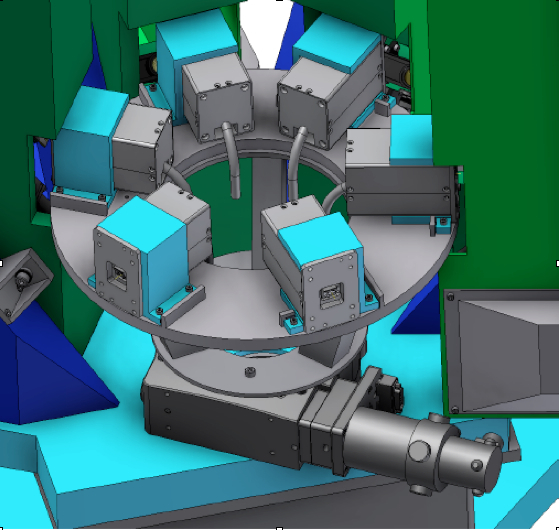}
\caption{Toggling mechanism for the selection of the observing mode.
}
\label{fig_toggling}
\end{figure}

The Front-End transports the beam received from the Coud\'e, once corrected for
atmospheric dispersion by the ADC, to the common focal plane where the pickups for the 
spectrograph fiber feeds are located. While performing such a beam
conditioning the Front-End applies pupil and field stabilization. They are
achieved via two independent control loops each composed of a technical camera
and a tip-tilt stage. Another dedicated stage delivers a focusing function. In
addition, the Front-End handles the injection of the calibration light, prepared
in the Calibration Unit, into the fibers and then into the spectrograph.  
As calibration sources we foresee a laser frequency comb (LFC), with as backup two ThAr lamps (one for simultaneous reference and one for calibration) and a  
calibration light is foreseen  to be composed of 2 ThAr, one for simultaneous
 Fabry-P\'erot unit. A top view of the Front-End arrangement is shown in Fig. \ref{label8}. A toggling mechanism shown in Fig. \ref{fig_toggling} handles the selection between the possible observational modes described below in a fully passive way.

The Fibre-Link subsystem relays the light from the Front-End to the spectrograph
and forms the spectrograph pseudo-slit inside the vacuum vessel. The 1-UT mode
uses a bundle of 2 octagonal fibres  each, 
one for the object and one for the sky or simultaneous reference. In the
high-resolution (singleHR) mode the fiber has a core of 140~$\mu$m, equivalent
to 1 arcsec on the sky; in the ultra-high resolution (singleUHR) mode the fibre
core is 70~$\mu$m and the covered field of view 0.5 arcsec.
The fibre entrances are organized in pickup heads that are moved to the focal plane of the Front End   when the specific bundle  of the specific mode is selected. 
In the 4-UT mode (multiMR) four object fibres and four sky/reference fibres are fed simultaneously by  the four telescopes. The four object fibres will finally feed a single square 280 $\mu$m object fibre, while the four sky/reference fibres will feed a single square 280 $\mu$m sky/reference fibre. Also in the 4-UT mode the spectrograph will {\it see} a pseudo slit of four fibre images, although they will be square and twice as wide as the 1-UT fibres. Another essential task performed by the Fibre-link subsystem is the light scrambling. The use of a double-scrambling optical system will ensure both scrambling of the near field and far field of the light beam. A high scrambling gain, which is crucial to obtain the required RV precision in the 1-UT mode is achieved by the use of octagonal fibres (Chazelas et al. 2011). 

\subsection{Observing modes}
\begin{figure*}
\center{\includegraphics[width=160mm]{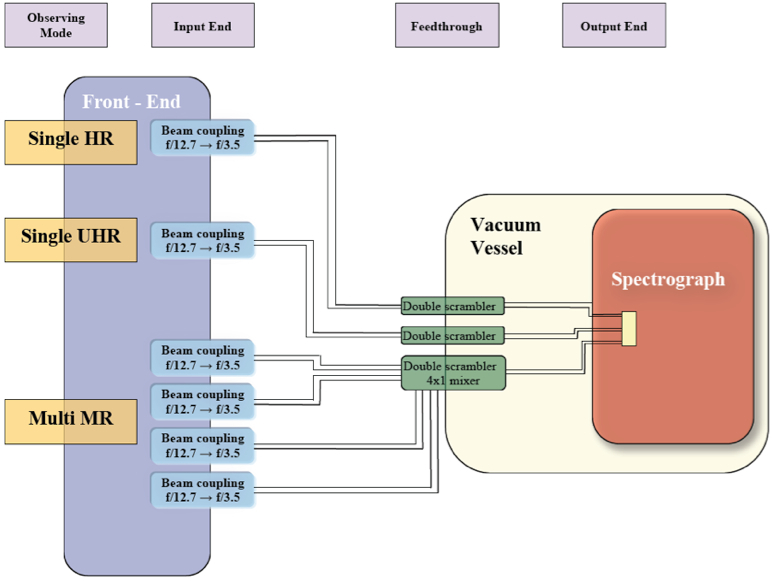}}
\caption{Scheme of the different observing modes.
}
\label{fig_modes}
\end{figure*}

The extreme precision  of ESPRESSO will be obtained by  improving well-known
HARPS concepts. The light of one or several UTs is fed by means of the front-end
unit into optical fibres that scramble the light and provide excellent
illumination stability to the spectrograph. In order to improve light
scrambling, non-circular fibre shapes will be used. The target fibre can be fed
either with the light from the astronomical object or one of the calibration sources.
The reference fibre will receive either sky light (faint source mode) or
calibration light (bright source mode). In the latter case -- the famous
simultaneous-reference technique adopted in HARPS -- it will be possible to
track instrumental drifts down to the \cms\ level. In this
mode the measurement is photon-noise limited and detector read-out noise
negligible. In the faint-source mode, instead, detector noise and sky background
may become significant. In this case, the second fibre will allow to measure the
sky background, while a slower read-out and high binning factor will reduce the
detector noise. As summarized   in Table 1, ESPRESSO will have three instrumental
modes: singleHR, singleUHR and multiMR. Each mode will be available with two
different detector read-out modes optimized for low and high-SNR measurements,
respectively. In high-SNR (high-precision) measurements the second fiber will be
fed with the simultaneous reference, while in the case of faint objects it shall
be preferred to feed the second fiber with sky light. A schematic view of the
different observing modes is shown in Fig. \ref{fig_modes}.

\subsection{Performances}

\begin{figure}
\includegraphics[width=83mm]{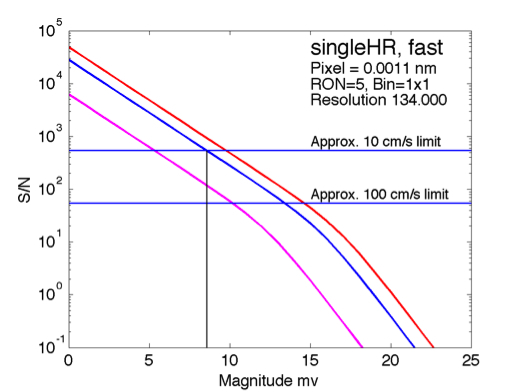}
\caption{Achievable signal-to-noise ratio as a function of stellar visible magnitude for the singleHR. Red, blue, and violet curves indicate exposure times of 3600 s, 1200 s, and 60 s, respectively.}
\label{label9a}
\end{figure}

\begin{figure}
\includegraphics[width=83mm]{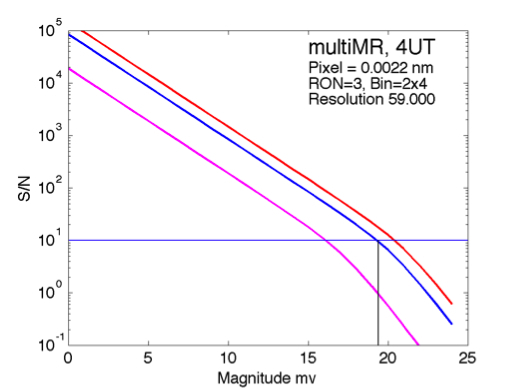}
\caption{  As Fig. \ref{label9a} but for the multiMR mode with binning of $2{\times} 4$ pixels.
}
\label{label9b}
\end{figure}

The observational efficiency of ESPRESSO is shown in Figs. \ref{label9a}  and  \ref{label9b}. In the singleHR, ${R\approx 134\,000}$ mode, a SNR = 10 per extracted pixel is obtained in 20 minutes on a 
${V = 16.3}$ star, or a SNR = 540 on a ${V = 8.6}$ star. We have estimated that at this resolution this SNR value will lead to 10 \cms\ RV precision for a non-rotating K5 star. For an F8 star, the same precision would be achieved for ${V = 8}$. In the multiMR mode, at ${R\approx 60\,000}$, a  SNR of $\approx$\,10  is achieved on a ${V = 19.4}$ star with a 
20 minute  exposure, a binning 2$\times$4, and a slow read-out of the CCD.

\begin{table}
\caption{ESPRESSO's observing modes.}
\centering
\begin{tabular}{lccc}
\hline\noalign{\smallskip}
Par./Mode &  HR (1UT)  & MR(4UTs)   &  UHR \\
\hline\noalign{\smallskip}
Wave. range  & 380--780 nm & 380--780 nm & 380--780 nm \\
Resol. Power & 134\,000& 59\,000 & $\approx$\,200\,000 \\
Aper. on Sky & 1\farcs0 & 4\arcsec$\times$1\arcsec & 0\farcs5 \\
Spec. Samp. & 4.5 pix & 11 pix   & 2.5 pix \\
Spat. Samp.  & 11$\times$2  pix & 22$\times$2 pix  & 5$\times$2 pix \\
Sim. Ref. & Yes (no sky) & Yes (no sky)& Yes (no sky) \\
Sky Sub.& Yes (no ref.)  & Yes (no ref.)&   Yes (no ref.) \\
Tot.  Eff. & 11\,\% & 11\,\%& 5\,\% \\
 \hline
\end{tabular}
\end{table}

\subsection{Design}
Several  {\it tricks} have been used to obtain high spectral resolution and efficiency despite the large size of the telescope and the 1  arcsec sky aperture of the instrument. 
In order to minimize the size of the optics, particularly of collimator and
\'echelle grating, ESPRESSO implements an anamorphic optics, the APSU, which
compresses the size of the pupil in the direction of the cross-dispersion. The
pupil is then sliced in two by a pupil slicer and the slices are overlapped on
the \'echelle grating, leading to a doubled spectrum on the detector. The shape
and size of both the pupil and the fibre image is shown in Fig. \ref{label3} for
various locations along the optical beam of the spectrograph. Without using this
method, the collimator beam size would have been 40 cm in diameter and the size
of the \'echelle grating would have reached a size of $240 {\times} 40$ cm. The
actual ESPRESSO design foresees the use of an Echelle grating of {\it only} 
$120 {\times} 20$ cm and of much smaller optics (collimators, cross dispersers, etc.). The \'echelle grating will be an  R4 Echelle of 31.6 l\,mm$^{-1}$ and a blaze angle of 76$^{\circ} $. This solution significantly reduces the overall costs. The drawback is that each spectral element will be covered by more detector pixels given the  doubled  image of the target fibre and its  elongated shape on the CCD. In order to avoid increased detector noise, heavy binning will be done in the case of faint-object observations, especially in the 4-UT mode.

 \begin{figure*}
\center{\includegraphics[width=160mm]{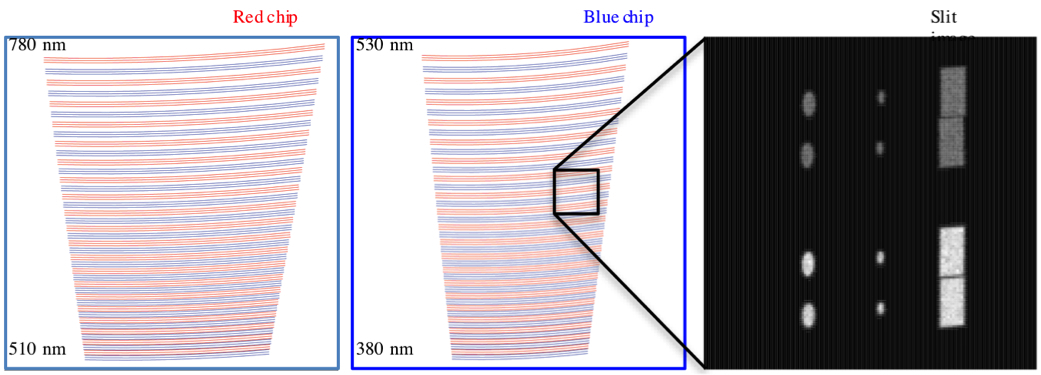}}
\caption{\emph{Left}: format of the red spectrum; \emph{middle}: format of the blue spectrum; \emph{right}: zoom on the pseudo slit. This latter shows the image of the target (bottom) and sky fiber (top). Each fiber is re-imaged in two slices. The three sets of fibers corresponding (from left to right) to the standard resolution 1-UT mode, ultra-high resolution 1-UT mode, and mid-resolution 4-UT mode, are shown simultaneously.
}
\label{label2}
\end{figure*}

\begin{figure}
\includegraphics[width=80mm]{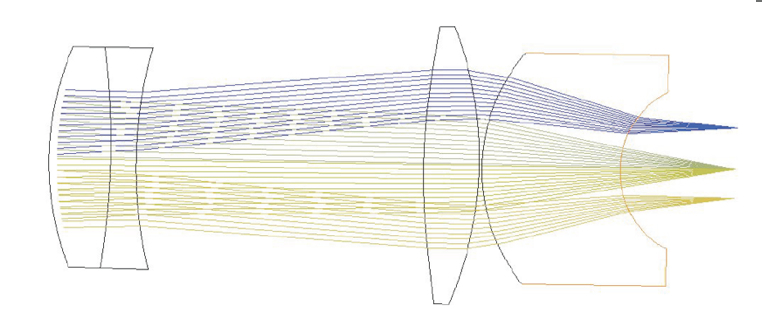}
\caption{Blue Camera elements. More than 80\,\% of  encircled energy measured at 9 wavelengths for all orders results within the CCD pixel size of 10 $\mu$m.
}
\label{fig_camera}
\end{figure}

\begin{itemize}

\item{  {\it The Anamorphic Pupil Slicing Unit (APSU).}   At the spectrograph entrance the APSU  shapes the beam in order to compress it in cross-dispersion and splits in two smaller beams, while superimposing them on the Echelle grating to minimize its size. The rectangular white pupil is then re-imaged and compressed.}

\item{ {\it Dichroic.}  Given the wide spectral range, a dichroic beam splitter separates the beam in a blue and a red arm, which in turn allows to optimizing each arm for image quality and optical efficiency. }

\item{ {\it Volume Phase Holographic Gratings (VPHGs).} The cross-disperser has the function of separating the dispersed spectrum in all its spectral orders. In addition, an anamorphism is re-introduced to make the pupil square and to compress the order height such that the inter-order space and the SNR per pixel are both maximized. Both functions are accomplished using Volume Phase Holographic Gratings (VPHGs) mounted on prisms.} 

\item{{\it Fast Cameras.}  Two optimised camera lens systems image the full
spectrum from 380 nm to 780 nm on two large $92{\times}92$ mm CCDs with 10-$\mu$m
pixels. The blue camera is shown in Fig. \ref{fig_camera} has an entrance
aperture of $150{\times} 150$  mm, a Focal length of 400 mm at 450 nm and a focal ratio of $F$/2.6.}

\end{itemize}

 A sketch of the optical layout is shown in Fig. \ref{label1}. The spectral
 format covered by the blue and the red chips as well as the shape of the pseudo
 slit are shown in Fig. \ref{label2}. The spectrograph is also equipped with an advanced exposure meter that measures the flux entering the spectrograph as a function of time. This function is necessary to compute the weighted mean time of exposure at which the precise relative Earth motion must be computed and corrected for in the RV measurement. Its innovative design (based on a simple diffraction grating) allows a flux measurement and an RV correction at different spectral channels, in order to cope with possible chromatic effects that could occur during the scientific exposures. The use of various channels also provides a redundant and thus more reliable evaluation of the mean time of exposure.

\subsection{The opto-mechanics}

\begin{figure}
\includegraphics[width=80mm]{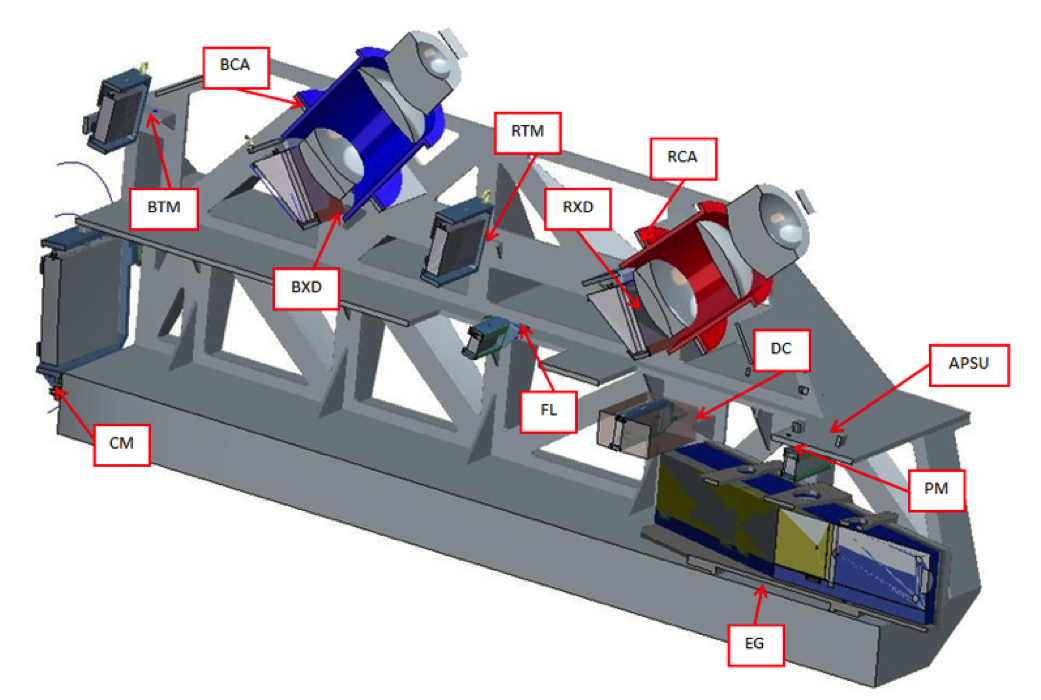}
\caption{Opto-mechanics of the ESPRESSO spectrograph (APSU: Anamorphic Pupil-Slicer Unit, BCA: Blue Camera, BTM: Blue Transfer Mirror, BXD: Blue Cross-Disperser, CM: Main Collimator, DC: Dichroic, EG: Echelle Grating, FL: Field Lens, PM: Field Mirror, RCA: Red Camera, RTM: Red Transfer Mirror, RXD: Red Cross-Disperser.
}
\label{label4}
\end{figure}

\begin{figure*}
\center{\includegraphics[width=160mm]{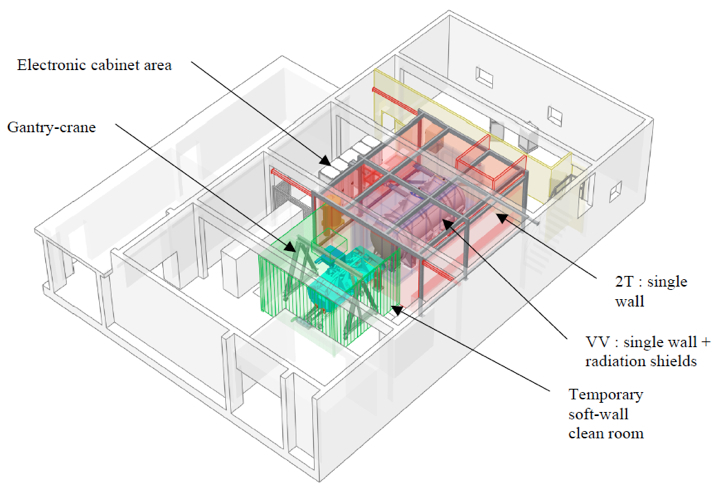}}
\caption{ESPRESSO inside the CCL, vacuum vessel, and multi-shell thermal control system.
}
\label{fig_vacuum}
\end{figure*}

ESPRESSO is designed to be an ultra-stable spectrograph capable of reaching RV
precisions of the order of 10 \cms, i.e. one order of magnitude better than its
predecessor HARPS. ESPRESSO is therefore designed with a totally fixed
configuration and for the highest thermo-mechanical stability. The spectrograph
optics are mounted in a 3-dimensional optical bench specifically designed to keep
the optical system within the thermo-mechanical tolerances required for
high-precision RV measurements. The bench is mounted in a vacuum vessel in which
10$^{-5}$ mbar class vacuum is maintained during the entire duty cycle of the
instrument. An overview of the opto-mechanics is shown in Fig. \ref{label4}. The
temperature at the level of the optical system is required to be stable at the
mK level in order to avoid both short-term drift and long-term mechanical
instabilities. Such an ambitious requirement is obtained by locating the
spectrograph in a multi-shell active thermal enclosure system  as shown in Fig. \ref{fig_vacuum}.  Each shell will improve the temperature stability by a factor 10, thus getting from typically Kelvin-level variations in the CCL  down to 0.001 K stability inside the vacuum vessel and on the optical bench.

\subsection{Large-area CCDs}

\begin{figure}
\includegraphics[width=80mm]{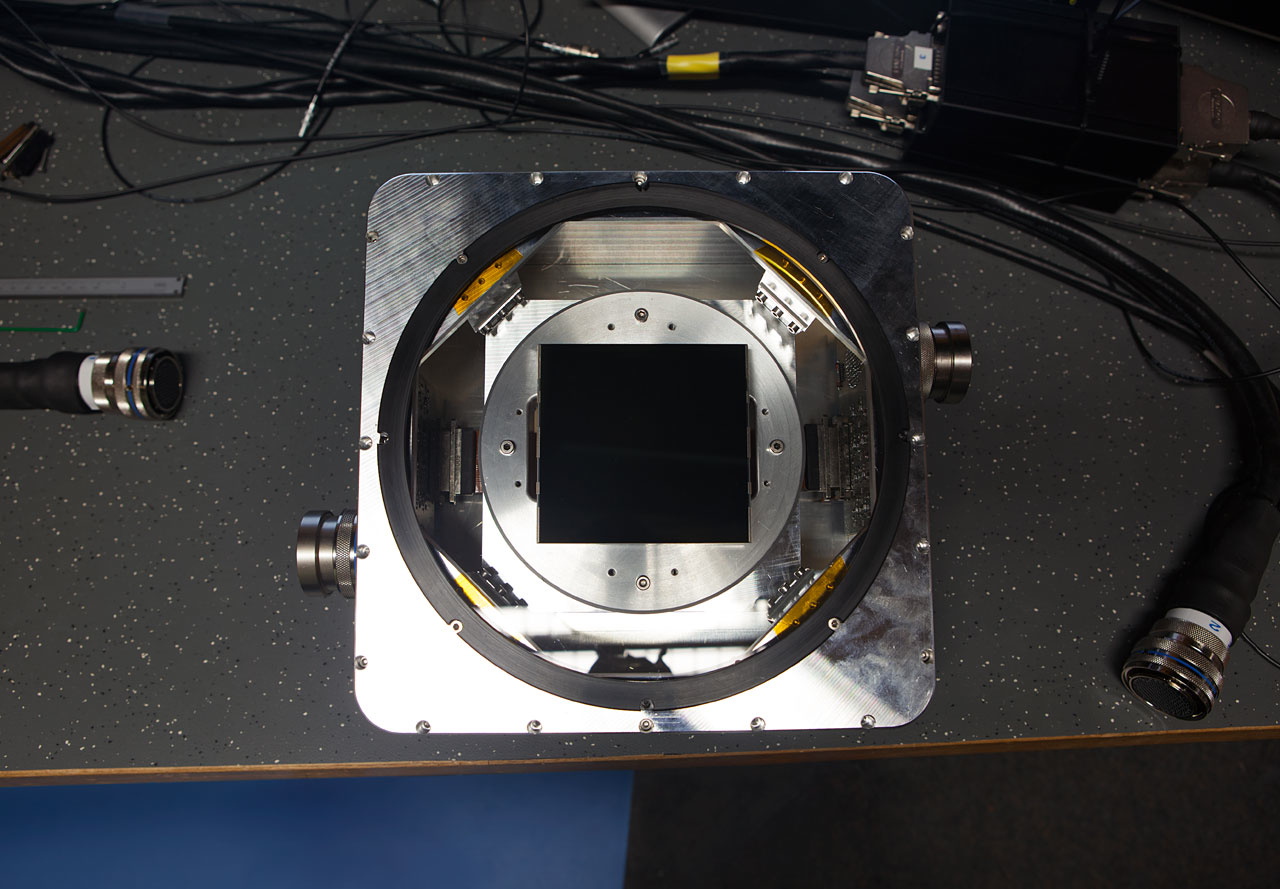}
\caption{  The first ESPRESSO e2v CCD   to be used in the cameras of the ESPRESSO instrument. These very large CCD samples  have more than 80 million pixels over an area 
92$\times$92 millimetres.
}
\label{fig_ccd}
\end{figure}

ESPRESSO presents also innovative solutions in the area of the CCDs, their
packages and cryostats. One of the world's largest monolithic state-of-the-art
CCDs was selected to properly utilize the optical field of ESPRESSO and to
further improve the stability compared to a mosaic solution, as employed
in HARPS. The sensitive area of the e2v chip is $92 {\times} 92$ mm covering 
8.46$\times  10^{7}$ pixels of 10 $\mu$m  size. Fast read out of such a large chip is achieved by using its 16 output ports  at high speed. Other requirements on CCDs are very demanding, e.g. in terms of Charge Transfer Efficiency (CTE) and all the other parameters affecting the definition of the pixel position, immediately reflected into the radial-velocity precision and accuracy. The CCDs are currently being procured by ESO from the e2v supplier. Firs  engineering devices have already been received and one is shown in Fig.~\ref{fig_ccd}. First warm technical light with the ESO custom-made components (NGC controller, cabling, cryostat electronics, firmware and mock-up mechanics) took place in July 2013. ESPRESSO's aimed precision of 10 cm s$^{-1}$  rms requires measuring spectral line position changes of 2 nm (physical) in the CCD plane, equivalent to only 4 times the silicon lattice constant! For better stability and thermal-expansion matching the CCD package is made of silicon carbide. The package of the CCDs, the surrounding mechanics and precision temperature control inside the cryostat head and its cooling system, as well as the thermal stability and the homogeneous dissipation of the heat locally produced in the CCDs during operation are of critical importance. ESO has built therefore a new {\it superstable} cryostat that has already demonstrated excellent short-term stability. A breadboard of the concept is currently being tested and the results will drive the design of the final ESPRESSO detector system.

\subsection{A laser frequency comb}

In order to track possible residual instrumental drifts, ESPRESSO will implement the simultaneous reference technique (similarly to HARPS; see, e.g., Baranne et al. 1996), i.e. the spectrum of a spectral reference will be recorded simultaneously on the scientific detector. Nevertheless, all types of spectrographs need to be wavelength-calibrated in order to assign to each detector pixel the correct wavelength with a repeatability of the order of $ { \Delta \lambda}/ { \lambda} \approx 10^{-10}$. A necessary condition for this step is the availability of a suitable spectral wavelength reference. None of the currently used spectral sources (thorium argon spectral lamps, iodine cells, etc.) would provide a spectrum sufficiently wide, rich, stable and uniform for this purpose. Therefore, the baseline source for the calibration and simultaneous reference adopted for ESPRESSO is a laser frequency comb. The LFC presents all the characteristics indispensable for a precise wavelength calibration and provides a link to the frequency standard. The procurement of an LFC suited for ESPRESSO is going on and  at the time of writing, placement of a contract for a LFC covering the full 380 - 760 nm range was imminent. In parallel, ESO has been developing such a source for HARPS, in collaboration with other institutes and industrial partners (Lo Curto et al. 2012). As a back-up solution and in order to minimize risks, also a stabilized Fabry-P\'erot is currently also under development within the consortium.

\section{ESPRESSO's data flow}

Following the very positive experience gained with HARPS, ESPRESSO has been conceived since the beginning not to be a {\it simple} standalone instrument, but actually a {\it science-generating machine}. The final goal is to provide the user with scientific data as complete and precise as possible in a short time (within minutes) after the end of an observation, increasing in this way the overall efficiency and the ESPRESSO scientific output. For this purpose a software-cycle integrated view, from the observation preparation through instrument operations and control to the data reduction and analysis has been adopted since the early phases of the project. Coupled with a careful design this will ensure optimal compatibility, easiness of operations and maintenance within the existing ESO Paranal Data Flow environment both in service and visitor mode. ESPRESSO Data Flow presents the following main subsystems:

 \begin{figure*}
\center{\includegraphics[width=156mm]{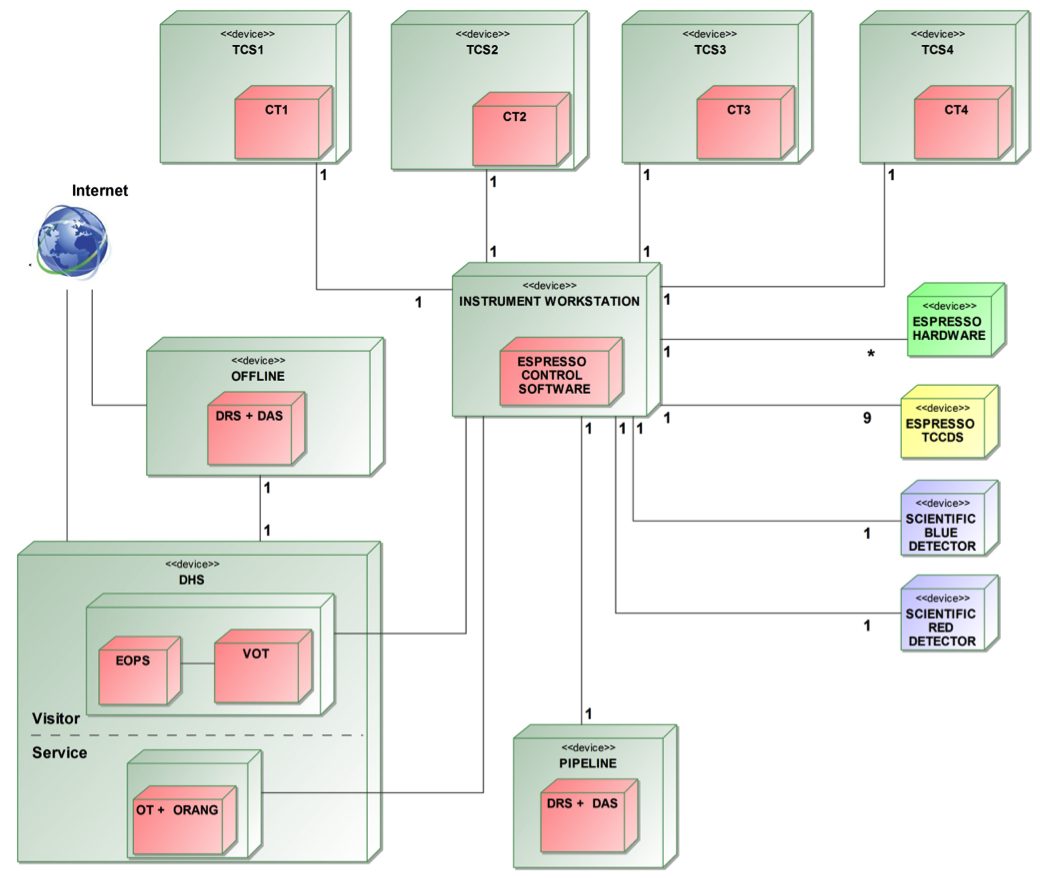}}
\caption{Overview of ESPRESSO's control system.  \vspace{-2mm}
}
\label{label10}
\end{figure*}

 \begin{itemize}
 
\item{\it  The ESPRESSO Observation Preparation Software (EOPS)}: a dedicated visitor tool (able to communicate directly with the VOT - Visitor Observing Tool) to help the observer to prepare and schedule ESPRESSO observations at the telescope according to the needs of planet-search surveys. The tool will allow users to choose the targets best suited for a given night and to adjust the observation parameters in order to obtain the best possible quality of data.

\item{\it The Data Analysis Software (DAS)}: dedicated data analysis software will allow to obtain the best scientific results from the observations directly at the telescope. A robust package of recipes  tailored to ESPRESSO, taking full advantage of the existing ESO tools (based on CPL and fully compatible with Reflex), will address the most important science cases for ESPRESSO by analyzing (as automatically as possible) stars and quasar spectra (among others, tasks will be performed such as line Voigt-profile  fitting, estimation of stellar atmospheric parameters, quasar continuum fitting, identification of absorption systems).

\item{\it The Data Reduction Software (DRS)}: ESPRESSO will have a fully automatic data reduction pipeline with the specific aim of delivering to the user high-quality reduced data, science ready, already in a short time after an observation has been performed. To this purpose the  computation of the RV at a precision better than 10 cm/s will be an integral part of the DRS. Coupled with the need to optimally remove the instrument signature, to take account the complex spectral and multi-HDU FITS format, the handling of the simultaneous reference technique and the multi-UT mode will make the DRS a truly challenging component of the DFS chain.

\item{\it Templates and control}: compared to other standalone instruments, the
main reason for ESPRESSO acquisition and observation templates complexity will
be the possible usage of any combination of UTs, besides the proper handling of
the simultaneous reference technique. Coupled with the fact that at the
instrument control level PLCs (Programmable Logical Controllers) and new COTS
(Component Off-The Shelf) TCCDs will be adopted instead of the (old) VME
technology, ESPRESSO will contribute to open the new path for the control
systems of future ESO instrumentation. A general overview of the ESPRESSO
control system is shown in Fig. \ref{label10}. 

\end{itemize}

\subsection{End-to-end operation}
 In the  singleHR mode  ESPRESSO can be fed by any of the four UTs, possibility which significantly improves the scheduling flexibility for ESPRESSO programmes and optimizes the use of VLT time in general. Scheduling flexibility is a fundamental advantage for survey programmes like RV searches for extrasolar planets or time-critical programmes like studies of transiting planets. The singleHR mode itself will thus greatly benefit from the implementation of the multiMR mode. The overall efficiency and the scientific output of long-lead programmes can be considerably increased if an integrated view of the operations is adopted. ESPRESSO shall not be considered as a stand-alone instrument but as a science-generating machine. Full integration of the data-flow system as described above is fundamental. ESPRESSO will deliver full-quality scientific data less than a minute after the end of an observation.

\acknowledgements

The ESPRESSO project is supported by the Swiss National Science Foundation through the FLARE funding Nr. 206720-137719 as well as by the Geneva University by providing infrastructures, human resources and direct project funding. 
We acknowledge the support from Funda\c{c}\~ao para a
Ci\^encia e a Tecnologia (FCT, Portugal) through FEDER funds in program COMPETE, as well as through national funds, in the form of grants reference PTDC/CTE-AST/120251/2010 (COMPETE reference FCOMP-
01-0124-FEDER-019884) and RECI/FIS-AST/0176/2012 (FCOMP-01-0124-FEDER-027493). We also acknowledge the support from the European Research Council/European Community under the FP7 through Starting Grant agreement number 239953. NCS also acknowledges the support in the form of a Investigador FCT contract funded by Funda\c{c}\~ao para a Ci\^encia e a Tecnologia (FCT) /MCTES (Portugal) and POPH/FSE (EC)
 The PI, on behalf of the ESPRESSO Executive Board,  warmly acknowledges  all the persons who have contributed and still contribute in a direct or indirect way to the ESPRESSO project.

\end{document}